\documentclass[reqno,11pt]{amsart} 
\usepackage{cite}
\theoremstyle{plain}
\newtheorem{Thm}{Theorem}%[chapter]
\newtheorem{Prop}[Thm]{Proposition}
\newtheorem{Lem}[Thm]{Lemma}
\theoremstyle{definition}
\newtheorem{Def}[Thm]{Definition}
\numberwithin{equation}{section}
\newcommand{\Bb}{\mathbb{R}}
\newcommand{\Bc}{\mathbb{C}}

\newcommand{\Bn}{\mathbb{N}}
\newcommand{\F}{{\mathcal F}}
\newcommand{\calH}{{\mathcal H}}
\newcommand{\calO}{{\mathcal O}}
\newcommand{\calC}{{\mathcal C}}
\newcommand{\calK}{{\mathcal K}}
\newcommand{\calS}{{\mathcal S}}
\newcommand{\Po}{P_+^{\uparrow}}
\newcommand{\Potild}{\tilde{P}_+^{\uparrow}}
\newcommand{\Lor}{L_+^{\uparrow}}
\newcommand{\He}{\calH^{(1)}}
\newcommand{\half}{{\frac{1}{2}}} 
\renewcommand{\d}{{\rm d}}
\newcommand{\unity}{1}
\newcommand{\utilde}[1]{\mathop{#1}\limits_{\widetilde{\phantom{\textstyle
 #1}}}}
\newcommand{\bfp}{{\boldsymbol{p}}}
\newcommand{\bfv}{{\boldsymbol{v}}}
\newcommand{\bfx}{{\boldsymbol{x}}}
\newcommand{\clo}{ {\mbox{\bf --}} }
 
\newcommand{\im}{{\rm Im }}
\newcommand{\supp}{{\rm supp}}
\newcommand{\eps}{\varepsilon}
\newcommand{\Om}{\Omega}
\newcommand{\Out}{{\rm out} }
\newcommand{\In}{{\rm in} }
\newcommand{\ex}{{\rm ex} }

\newcommand{\Sec}{\Sigma} % Set of charges
\renewcommand{\sec}{\alpha}  
\newcommand{\Sece}{\Sec^{(1)}} % Set of elementary charges
\newcommand{\Stom}{S_{\rm Tom}(W_1)}
\newcommand{\Stome}{S_{\rm Tom}} %^{(1)}
\newcommand{\Sgeo}{S_{\rm geo}} 
\newcommand{\Sgeoe}{S_{\rm geo}}  %^{(1)}
\newcommand{\DWR}{\Delta_{W_1}}  %{\Delta_{W_R}}
\newcommand{\JWR}{J_{W_1}}  %{J_{W_R}}

\newcommand{\Hsec}{\calH_{\sec}}
\newcommand{\Hbarsec}{\calH_{\bar\sec}}
\newcommand{\Esec}{E_{\sec}}

\newcommand{\msec}{{m_{\sec}}}
\newcommand{\ssec}{{s_{\sec}}}
\newcommand{\Hesec}{\Ee\Hsec} %{\calH_{\sec}^{(1)}}
\newcommand{\Hebarsec}{\Ee\Hbarsec} %{\calH_{\bar\sec}^{(1)}}
 
\newcommand{\ptilde}{\tilde{p}} 
\newcommand{\Ue}{U^{(1)}} 
\newcommand{\Uej}{U^{(1)}(j)} 
\newcommand{\Ums}{U_{m,s}}  %%%%  
\newcommand{\Umssec}{U_{\msec,\ssec}}  %%%% 
\newcommand{\Hms}{\calH_{m,s}}  %%%%  
\newcommand{\Ee}{E^{(1)}} 
\newcommand{\Emsec}{E_{\{\msec\}}}  
\newcommand{\Eeseck}{\Ee_{\sec,k}}

\newcommand{\Cop}{C}        % Charge conjugation operator 
\newcommand{\CPTop}{\Theta} % CPT operator  
\newcommand{\cone}{\calC}   %spacelike cone  

\title%[Bisognano-Wichmann Theorem]%
{The Bisognano-Wichmann Theorem for Massive Theories}
\author{Jens Mund}
\address{Institut f\"ur theoretische Physik, Universit\"at G\"ottingen, 
Bunsenstr.\ 9, 37\,073 G\"ottingen, Germany} 
\email{mund@theorie.physik.uni-goettingen.de} 
   \date{January 2001}
\begin{document} 
\begin{abstract}
The geometric action of modular groups for wedge regions 
(Bisognano--Wich\-mann property) 
is derived from the principles of local quantum physics 
for a large class of Poincar\'e covariant models in $d=4$. 
As a consequence, the CPT theorem holds for this class. 
The models must have a complete interpretation in terms of massive particles.  
The corresponding charges need not be localizable in 
compact regions: The most general case is admitted, namely 
localization in spacelike cones. 
\end{abstract}
\maketitle 
\section*{Introduction.} \label{secIntro} 
In local relativistic quantum theory~\cite{H96}, a model is specified
in terms of a net of local 
observable algebras and a representation of the Poincar\'e group,
under which the net is covariant. The Bisognano-Wichmann 
theorem~\cite{BiWi,BiWi2}  intimately connects 
these two, algebraic and geometric, aspects. 
Namely, it asserts that under certain conditions {\em modular
  covariance} is satisfied: 
The modular unitary group of the observable algebra 
associated to a wedge region coincides with the unitary group representing the 
boosts which preserve the wedge.  
Since the boosts associated to all wedge regions generate the
Poincar\'e group, modular covariance implies that the representation
of the Poincar\'e group is  encoded intrinsically in the net of local 
algebras. 
It has further important consequences: It implies the spin-statistics 
theorem~\cite{Kuck,GL} and, as Guido and Longo
have shown~\cite{GL}, the CPT theorem. It 
also implies essential Haag duality, which is an important input to 
the structural analysis of  charge superselection sectors~\cite{DHRIII,DHRIV}. 

Counterexamples~\cite{Yng94,BGL93,BDFS} demonstrate that modular covariance 
does not follow from the basic principles of quantum field theory 
without further input. But its remarkable implications assign a
significant role to this property, and it is 
desirable to find physically transparent conditions under which it
holds. 
Bi\-sognano and Wichmann have shown modular covariance 
to hold in theories where the field algebras are generated by finite-component 
Wightman fields~\cite{BiWi,BiWi2}. 
In the framework of algebraic quantum field theory,
Borchers  has shown
that the mo\-dular objects associated to wedges have the
correct\footnote{namely, as expected from modular covariance} 
commutation relations with the translation operators as a consequence
of the positive energy requirement~\cite{Borch92}. Based on his 
result, Brunetti, Guido and Longo derived modular covariance 
for conformally covariant theories~\cite{BGL93}. 
In the Poincar\'e covariant case, sufficient conditions for modular
covariance of technical nature have been found by several 
authors~\cite{Borch98,BY00,SHW98,Kuck00,GL00} 
(see~\cite{BY00} for a review of these results). 

In the present article, we 
%show that modular covariance, and hence the CPT theorem, can be derived 
derive modular covariance in the setting of local quantum physics
for a large class of massive models.  
Specifically, the models must contain massive particles whose scattering states
span the whole Hilbert space (asymptotic completeness). 
Further, within each  
charge sector the occurring particle masses must be isolated
eigenvalues of the mass operator. 
The corresponding representation of 
the covering group of the Poincar\'e group must have no accidental
degeneracies; {\it i.e.\ }for each mass and charge there is one single
particle multiplet under the gauge group (the group of inner symmetries). 
We admit the most general localization properties for
the charges carried by these particles, namely localization in
spacelike cones~\cite{BuF}. 

A byproduct of our analysis is that the CPT theorem holds
under these rather general and transparent conditions. 
It must be mentioned that 
Epstein has already proved a rudimentary version of the CPT theorem for
massive theories in the framework of local 
quantum physics~\cite{Ep}. But it refers only to the S-matrix (and not 
to the local fields), and is derived only for compactly localized charges. 
It must also be  mentioned that the spin-statistics theorem, 
which is a consequence of modular covariance and needs not be assumed
for our derivation, 
%not necessary for our derivation, but  a consequence of modular covariance 
has already been proved by Buchholz and Epstein~\cite{BuEp} for massive
theories with  charges localizable in spacelike cones.   
%In fact, the present analysis is an extension of their proof. 

The article is organized as follows. In Section~\ref{secAssRes}, the 
general framework is set up and our assumptions concerning 
the particle spectrum are made precise, as well as our notion of 
modular covariance. We state our main result 
in Theorem~\ref{BiWi}. 
The proof will proceed in two steps: In Section~\ref{secH1}, 
single-particle versions of the Bisognano-Wichmann and the CPT theorems 
are derived (Theorem~\ref{ThmModCov1}). This is an extension of 
Buchholz and Epstein's proof ~\cite{BuEp} of the spin-statistics theorem for 
topological charges. In Section~\ref{secHex} we prove modular 
covariance via Haag-Ruelle scattering theory 
(Proposition~\ref{ModCovScat}). As mentioned, this already implies 
the CPT theorem~\cite{GL}. Yet for the sake of 
self consistency, we show in Section~\ref{secCPT} that the 
CPT theorem can be derived directly from our assumptions via 
scattering theory (Proposition~\ref{CPT}). 
\section{Assumptions and Result.}  \label{secAssRes} 
In the algebraic framework, the fundamental objects of a quantum field
theory are the observable algebra and the  physically relevant 
representations of it. The set of  equivalence classes of these 
representations, or charge superselection sectors, has the structure
of a semigroup. 
We will assume that it is generated by a set of 
``elementary charges'' which correspond to massive particles. 
Then all relevant charges are localizable in spacelike
cones~\cite{BuF}. Under these circumstances and if Haag duality holds, 
it is known~\cite{DR90} that the theory may be equivalently described
by an algebra of (unobservable) charged field operators localized in
spacelike cones, and a compact gauge group acting on the fields. 
The observable algebra is then the set of gauge invariant elements of 
the field algebra, and modular covariance of the former
is equivalent to modular covariance of the latter~\cite{Kuck,GL}. 
We take the field algebra framework as the starting point of our
analysis. It is noteworthy that then essential Haag duality 
needs not be assumed for our result, but rather follows from it. 

Let us briefly sketch this framework. 
The Hilbert space   
$\calH$ carries a unitary representation $U$ of the universal covering group 
$\Potild$ of the Poincar\'e group which has positive energy, {\it
  i.e.\ }the joint spectrum of the generators $P_\mu$ of the translations is
contained in the closed forward lightcone. There is a unique, up to a
factor,  invariant vacuum vector $\Omega.$  Further, there is a
compact group $G$ (the gauge group) of unitary operators on $\calH$
which commute with the representation $U$ and leave $\Omega$
invariant. 

For every spacelike cone\footnote{a spacelike cone is a region in
  Minkowski space of the form $\cone=a+\cup_{\lambda>0}\lambda \calO,$ where
  $a\in\Bb^4$ is the apex of $\cone$ and $\calO$ is a double cone whose
  closure does not contain the origin.} 
$\cone$ there is a von Neumann algebra $\F(\cone)$ of so-called field 
operators acting in $\calH.$ 
The family $\cone\rightarrow\F(\cone),$
together with the representation $U$ and the group $G,$ 
satisfies the following properties. 

 0) {\em Inner symmetry:} For all $\cone$ and all $V\in G$ 
\[ V\,\F(\cone)\,V^{-1} =\F(\cone)\,.   \]

 i) {\em Isotony:} $\cone_1\subset \cone_2$ implies 
 $\F(\cone_1)\subset \F(\cone_2).$  

 ii) {\em Covariance:} For all  $\cone$ and all $g\in\Potild$ 
\begin{equation*} % \label{eqUgeo}
  U(g)\,\F(\cone)\,U(g)^{-1}=\F(g\,\cone) \,.  
\end{equation*}

 iii) {\em Twisted locality:} 
There is a Bose-Fermi operator $\kappa$ in the center of $G$ %$\in G\cap G'$ 
with $\kappa^2=1,$ determining the spacelike commutation relations of fields: 
let $ Z\doteq\frac{1+i\kappa}{1+i}.$ Then 
\begin{equation*} %\label{eqTwiLoc}
 Z\F(\cone_1)Z^* \subset \F(\cone_2)'\, 
\end{equation*}
if $\cone_1$ and $\cone_2$ are spacelike separated. 

 iv) {\em Reeh-Schlieder property:} For every $\cone,$ 
 $\F(\cone)\,\Omega$ is dense in $\calH.$ 

 v) {\em Irreducibility:} $\bigcap_\cone \F(\cone)' =\Bc\cdot \unity.$ 

Note that twisted locality (iii) is equivalent to normal commutation
relations~\cite{DHRI}: Two field operators which are localized in
causally disjoint cones anticommute if both operators are odd under
the adjoint action of $\kappa$ (fermionic) and commute if one 
of them is even (bosonic) and the other one is even or odd.  

Let 
\[ \calH = \bigoplus_{\sec\in\Sec} \Hsec
\]
be the factorial decomposition of $G'',$ %the defining representation of $G,$
with $\Sec$ the set of equivalence classes of irreducible unitary
representations of $G$ contained in the defining 
representation~\footnote{In fact, $\Sec$ contains all irreducible 
  representations of $G,$ and is in 1--1 correspondence with
  the superselection sectors of the observable algebra~\cite{DHRI}.}.  
The subspaces $\Hsec$ will be referred to as (charge) sectors, and two 
sectors corresponding to conjugate representations $\sec,\bar\sec$ of $G$ 
will be called conjugate sectors. We denote by $\Esec$ the projection
in $\calH$ onto $\Hsec,$ and by $d_\sec$  the (finite) dimension of the class 
$\sec.$ Note that the Poincar\'e representation commutes with $\Esec.$ 
Let $\Sece\subset \Sec$ be the set of charges carried by the 
particle types of the theory: $\sec\in\Sece$ if, and only if, the 
restriction of the mass  operator $\sqrt{P^2}$ to $\Hsec$ 
has non--zero eigenvalues. 

 vi) {\em Massive particle spectrum.} There are no massless particles,
 {\it i.e.\ }no eigenvectors of the mass operator with eigenvalue zero
   apart from the vacuum vector. For each $\sec\in\Sece,$ there
 is exactly one\footnote{Our results still hold if no restriction is
   imposed on the number of (isolated) mass values in each sector.}  
non--zero eigenvalue $\msec$ of $\sqrt{P^2}\Esec$, which is in 
addition isolated.  Further, the 
corresponding sub\-re\-presentation of $\Potild$ contains only one
irreducible representation, with multiplicity equal to 
$d_\sec.$ Thus, for each $\sec\in\Sece,$ there is one multiplet under $G$ of 
particle types with the same charge $\sec,$ mass and spin. 

 vii) {\em Asymptotic completeness:} The scattering states span the
 whole Hilbert space (see  equation~\eqref{eqHex}). 

The property of modular covariance, which we are going to derive from
these assumptions, means the following. 
Due to the Reeh-Schlieder property and locality, for every spacelike 
cone $\cone$ there is a Tomita operator~\cite{BraRob} $S_{\rm Tom}(\cone)$ 
associated to  $\F(\cone)$ and $\Omega:$ 
It is the  closed antilinear involution satisfying 
\begin{align*} % \label{eqStom}
S_{\rm Tom}(\cone)\,B\Omega&=B^*\Omega\quad\mbox{ for all }B\in\F(\cone)\,.\\
\intertext{Its polar decomposition is denoted as }
S_{\rm Tom}(\cone) &= J_{\cone}\,\Delta_{\cone}^{\half}\,.
\end{align*} 
The anti-unitary involution $J_{\cone}$ in this decomposition is called 
the {\em modular conjugation}, and the positive operator $\Delta_{\cone}$ 
gives rise to the so-called {\em modular unitary group} $\Delta_{\cone}^{it}$
associated to $\cone.$ 
Modular covariance means, generally speaking, that the Tomita
operators associated to a distinguished class of space-time regions 
have geometrical significance. This class is the set of wedge
regions, 
{\it i.e.\ }Poincar\'e transforms of the standard wedge region 
$W_1:$~\footnote{Wedges $W$ will be considered as limiting cases of 
spacelike cones. $\F(W)$ is the von Neumann algebra 
generated by all $\F(\cone)$ with $\cone\subset  W.$} 
\begin{align*} 
W_1 &\doteq\{\,x\in\Bb^4:\,|x^0|<x^1\;\}\,, 
\end{align*} 
and the geometrical significance is as follows. 
Let $\Lambda_1(t)$ denote the Lorentz boost in 
$x^1$-direction, acting on the coordinates $x^0,x^1$ as 
\begin{equation*}
\left( \begin{array}{cc}
 \cosh(t) &  \sinh(t)  \\
 \sinh(t) &  \cosh(t)
 \end{array} \right),
\end{equation*} %$\cosh t \,\unity +\sinh t\,\sigma_1$
and $\lambda_1(t)$ its lift to the covering group $\Potild.$ 
\begin{Def}
A theory is said to to satisfy {\em modular covariance}
if 
\begin{equation} \label{eqModCov}
  \DWR^{it} = U(\lambda_1(-2\pi t))\,. 
\end{equation}
\end{Def}
Other notions of modular covariance have been proposed in the literature 
(see, {\it e.g.}~\cite{Dav}), but this is the strongest one. In particular, it 
implies~\cite{GL} that the modular conjugation $\JWR$ 
has the geometric significance of representing the reflexion $j$ at
the edge of $W_1,$ which inverts the sign of $x^0$ and $x^1$ and acts
trivial on $x^2,x^3.$ More precisely, Guido and Longo have shown
in~\cite{GL} that equation~\eqref{eqModCov} implies that the operator 
$\CPTop\doteq Z^*\,\JWR$  acts geometrically correctly, {\it i.e.\ }satisfies 
\begin{align}
    \CPTop\,\F(W)\,\CPTop^{-1}&=\F(j W)\,\quad \label{eqjWj} 
%\mbox{ for     all }\cone\,.&& 
\end{align} 
for all wedge regions $W,$ and has the representation properties  
\begin{align} \label{eqUjgj}
\CPTop\,U(g)\,\CPTop^{-1}=U(jgj) \;\;   \mbox{ for
      all } g\in\Potild\,, \quad \CPTop^2=\unity\,.  
\end{align} 
Here we have denoted by $g\mapsto jgj$ the unique lift of the adjoint
action of $j$ on the Poincar\'e group to an automorphism of the 
covering group~\cite{Var2}. Since $\CPTop$ also sends each sector to its 
conjugate, equations~\eqref{eqjWj} and \eqref{eqUjgj}
exhibit $\CPTop$ as a CPT operator\footnote{Here we consider $j$ as the $PT$ 
transformation. The total space-time inversion arises from
$j$ through a $\pi$-rotation about the $1$-axis, and is thus also a 
symmetry (if combined with charge conjugation $C$). In odd-dimensional 
space-time, $j$ is the proper candidate for a 
symmetry in combination with $C,$ while the total space-time 
inversion is not. }. 
Thus, modular covariance~\eqref{eqModCov} implies the CPT
theorem. Further, the last two equations imply, by the Tomita-Takesaki 
theorem, that the theory satisfies {\em twisted Haag duality for wedges}, 
{\em i.e.\ } 
\begin{equation} \label{eqTHD}
 Z\F(W')Z^* = \F(W)'\,.
\end{equation}
Our main result is the following theorem. 
\begin{Thm} \label{BiWi} 
Let the assumptions 0) $,\ldots,$ vii) be satisfied. Then modular 
covariance, 
the CPT theorem as expressed by equations~\eqref{eqjWj} and
\eqref{eqUjgj}, and twisted Haag duality for wedges  hold. 
\end{Thm}
It is noteworthy that equation~\eqref{eqjWj} holds not only for wedge
regions, but also for spacelike cones if one replaces the family $\F$ with the
so-called dual family $\F^d.$ Namely, twisted Haag duality for wedge
regions implies that the dual family $\F^d(\cone)\doteq\cap_{W\supset
  \cone}\F(W)$ is still local. On this family, $\CPTop$ acts 
geometrically correctly, {\it i.e.\ }equation~\eqref{eqjWj} holds 
for all $\F^d(\cone)$~\cite{GL}.  

Our proof of the theorem will proceed in two steps: In the next 
section, modular covariance is shown to hold in restriction to the
single particle space (Theorem~\ref{ThmModCov1}). 
In Section~\ref{secHex} we show that modular covariance extends to 
the space of scattering states (Proposition~\ref{ModCovScat}). 
By the assumption of asymptotic
completeness this space coincides with $\calH,$ hence modular covariance
holds, implying the CPT theorem. 
\section{Modular Covariance on the Single Particle \\ Space.}  \label{secH1} 
As a first step, we prove single-particle versions of the 
Bisognano-Wichmann and the CPT theorems.  Let $E_I$, $I\subset\Bb,$ be
the spectral projections of the mass 
operator $\sqrt{P^2}.$ We denote by $\Ee$ the sum of all $\Emsec,$ where 
$\sec$ runs through the set $\Sece$ of single particle charges   and 
$m_\sec$ are the corresponding particle masses. The range of $\Ee$ is
called the  single particle space. % and denoted by $\He$. 

An essential step towards the Bisognano-Wichmann theorem is the 
mentioned result of
Borchers~\cite{Borch92,Borch93}, namely that the modular unitary group  
and the modular conjugation 
associated to the wedge $W_1$ have the correct commutation relations
with the translations. In particular, they commute with the mass
operator, which implies that $\Stom$ commutes with $\Ee.$ Let us denote the 
corresponding restriction by 
\begin{equation*} % \label{eqS1tom}
 \Stome\doteq \Stom\,\Ee \,. 
\end{equation*} 
Similarly, the representation $U(\Potild)$  leaves $\Ee\calH$ invariant,
giving rise to the subrepresentation 
\begin{equation*} % \label{eqU1}
  \Ue(g)\doteq U(g)\,\Ee\,, %\quad\mbox{ for } g\in\Potild\;, 
\end{equation*}    %$U(g)\,\Ee$ 
and one may ask if modular covariance holds on $\Ee\calH.$ We
show in this section that this is indeed the case, the line of
argument being as follows. Let $K$ denote the generator of the unitary
group of 1-boosts, $U(\lambda_1(t))=e^{itK}.$ We exhibit an anti-unitary 
``PT-operator'' $\Uej$ representing the reflexion $j$ on $\Ee\calH,$ and 
show that  $\Stome$ coincides with the ``geometric'' involution 
\begin{equation} \label{eqSgeo}
\Sgeoe\doteq \Uej\;e^{-\pi K}\,\Ee\, 
\end{equation} 
up to a unitary ``charge conjugation'' operator which commutes
with the representation $\Ue$ of $\Potild.$ 
By uniqueness of the polar decomposition, this will imply modular
covariance on $\Ee\calH.$ %((and J=CPT...))

We begin by exploiting our knowledge about $\Ue(\Potild).$ By assumption, for
each $\sec\in\Sece$ the subrepresentation $U(g)\Ee\Esec$  
contains only one equivalence class of irreducible representations. As
is well--known, the  latter is  fixed by the mass $\msec$ and a number 
$\ssec\in\half\Bn_0,$ the spin of the corresponding 
particle species.  

We briefly recall the so-called covariant irreducible representation 
$\Ums$ for mass $m>0$ and spin $s\in\half\Bn_0.$ 
The universal covering group of the proper ortho\-chronous Lorentz group 
$\Lor$ is  identified with $SL(2,\Bc)$~\cite{SW}. Explicitly, 
the boosts $\Lambda_k(\cdot)$ in $k$-direction and  
rotations $R_k(\cdot)$ about the $k$-axis, $k=1,2,3,$ lift to 
\begin{equation} \label{eqBooRot}
\lambda_k(t)= e^{\frac{1}{2}t\,\sigma_k} \quad \text{ and }\quad 
r_k(\omega)= e^{\frac{i}{2}\omega\,\sigma_k}\,,\quad\,k=1,2,3 \;,   
\end{equation}
respectively. 
The universal covering group $\Potild$ of the proper ortho\-chronous 
Poin\-car\'e group $\Po$ is the semidirect product of $SL(2,\Bc)$ with
the translation subgroup $\Bb^4,$ %on which $SL(2,\Bc)$ acts via $\Lambda.$  
elements being denoted by $g=(x,A).$ 
The representation $\Ums$ of $\Potild$ for $m>0$ and $s\in\half\Bn_0$
acts on a Hilbert space $\Hms$ of functions from the positive
mass  shell $H_m$ into $\Bc^{2s+1}.$ The latter, viewed as the space of 
covariant spinors of rank $2s,$ is acted upon by an irreducible representation
$V_s$ of $SL(2,\Bc)$ satisfying 
\[V_s(A^*)=V_s(A)^* \quad\mbox{ and } \quad
V_s(\bar{A})=\overline{V_s(A)}\,.\]
Let $\langle\;,\;\rangle$ denote the scalar product in $\Bc^{2s+1},$
and $\d \mu(p)$ the Lorentz invariant measure on the mass shell $H_m,$
and let, for $p=(p^0,\bfp)\in\Bb^4,$
\begin{align*}
\tilde{p}\doteq p^0\unity - \bfp \cdot \boldsymbol{\sigma} 
\quad \mbox{ and  } \quad 
\utilde{p}\doteq p^0\unity + \bfp \cdot \boldsymbol{\sigma} \;, 
\end{align*}
where $\boldsymbol{\sigma}=(\sigma_1,\sigma_2,\sigma_3)$ are the Pauli 
matrices. 
Then the scalar product in $\Hms$ is defined as 
\begin{equation} \label{eqScalProd}
 (\,\psi_1,\psi_2\,) = \int\d\mu(p)\;
\langle\,\psi_1(p)\,,\,V_{s}(\frac{1}{m}\ptilde)\,\psi_2(p)\, \rangle\,.  
\end{equation}
$\Ums$ acts on $\Hms$  according to 
\begin{equation}  \label{eqUms}
 \big(\Ums(x,A)\psi\big)(p) =  \exp(i x\cdot p) \; 
 V_{s}(A)\, \psi(\Lambda(A^{-1})\, p)\;, 
\end{equation} 
where $\Lambda:SL(2,\Bc)\rightarrow \Lor$ denotes the covering homomorphism. 
To this representation an anti--unitary  operator $\Ums(j)$ can be 
adjoined satisfying the representation properties 
\begin{equation} \label{eqUjgj1}
  \Ums(j)^2=\unity\;\quad\text{ and }\quad 
  \Ums(j)\,\Ums(g)\,\Ums(j)=\Ums(jgj) 
\end{equation}
for all  $g\in\Potild.$ Namely, it is given 
by~\footnote{A proof, as well as explicit formulae for the
  relevant group relations $jgj,$ are given in the Appendix for the 
convenience of the reader.} 
\begin{equation} \label{eqU1j} 
\big(\Ums(j)\psi\big)(p)\doteq V_s\big(\frac{1}{m}\utilde{p} 
\, \sigma_3\big)\,  \overline{\psi({-j}\, p)}\,.
\end{equation} 

By our assumption vi), we may identify the subrepresentation 
$U(g)\Ee\Esec$ with the direct sum of $d_\sec$ copies of 
$U_{m_\sec,s_\sec}(g).$ Then there are mutually orthogonal projections
$\Eeseck\subset \Esec,$ $k=1,\ldots d_\sec,$ in $\calH$ onto irreducible
subspaces such that 
\begin{align} 
 \Ee\,\Esec & =\sum_{k=1}^{d_\sec}\Eeseck\,, \label{eqEeseck}\\
 U(g)\,\Eeseck & = \Umssec(g) \,\Eeseck\quad\text{for
     all } g\in\Potild\,.  \nonumber 
\intertext{We define a ``PT--operator'' $\Uej$ on $\Ee\calH$ as the
  anti--linear extension of} 
\Uej \,\Eeseck &\doteq \Ums(j)\,\Eeseck. \nonumber 
\end{align}
Note that this definition of $\Uej$ depends on the choice of the 
decomposition~\eqref{eqEeseck}. 

We define now a closed antilinear operator $\Sgeoe$ in terms of the 
representation $\Ue,$ as anticipated, by equation~\eqref{eqSgeo}. 
Note that the group relation $j\,\lambda_1(t) \,j=\lambda_1(t)$ implies that
$\Sgeoe$ is, like $\Stome,$ an involution: it leaves its domain
invariant and satisfies $(\Sgeoe)^2\subset\unity.$ 
The following proposition is a corollary  of the 
article~\cite{BuEp} of Buchholz and Epstein. 
\begin{Prop}  \label{PropBuEp}
There is a unitary ``charge conjugation'' operator $\Cop$ on $\Ee\calH$ 
satisfying 
\begin{equation} \label{eqCop}
  \Cop E_\sec \,\Ee = E_{\bar\sec} \Cop \,\Ee \quad \text{and}\quad
  [\Cop,U(g)]\,\Ee=0\;\text{ for all }g\in\Potild,
\end{equation}
such that 
\begin{equation} \label{eqCSS}
  \Cop \, \Sgeo =\Stome\,.
\end{equation}
\end{Prop}
\begin{proof}
Let $\sec\in\Sece.$ Corresponding to the decomposition~\eqref{eqEeseck} 
of the particle multiplet $\sec$ into particle types $(\sec,k)$ there
is, for each $k$ in $\{1,\ldots, d_\sec\},$ a family of linear subspaces 
$\cone\rightarrow \F_{\sec,k}(\cone)\subset \F(\cone)$ 
%on which $U(\Potild)$ acts geometrically correct in the sense of
%equation~\eqref{eqUgeo}, 
satisfying 
\begin{align} \label{eqFseck}
 \Ee\,\F_{\sec,k}(\cone)\,\Omega&=\Eeseck\,\F(\cone)\,\Omega\,,
\end{align} 
see {\it e.g.\ }~\cite{DK,DHRI}. Note that the closures of the above
vector spaces are independent of $\cone$ by the Reeh--Schlieder
property and span $\Hesec$ if $k$ runs through $\{1,\ldots,d_\sec\}.$ 
Similarly, the ``anti--particle'' Hilbert spaces 
\begin{equation}  \label{eqAntiPart}
\Ee\,\F_{\sec,k}(\cone)^*\,\Omega^\clo\;
\end{equation} 
are independent of $\cone,$ mutually orthogonal for different $k,$ and
span $\Hebarsec$ if $k$ runs through $\{1,\ldots,d_\sec\}$ 
(note that $d_{\bar\sec}=d_\sec$).  
Buchholz and Epstein~\cite{BuEp} have shown the particle --
anti-particle symmetry in this situation: 
For each $\sec\in\Sece$ and $k=1,\ldots,d_\sec$ there is a
unitary map $\Cop_{\sec,k}$ from the closure $\Eeseck\,\calH$  of the vector 
space~\eqref{eqFseck} onto 
the  space~\eqref{eqAntiPart} intertwining the respective 
(irreducible) subrepresentations of $\Potild$. 
We  now  recall in detail the relevant result of Buchholz and Epstein. 
Denote by $\F_{\sec,k}^\infty(\cone)$ the set of
field operators $B\in\F_{\sec,k}(\cone)$ such that the map 
$g\mapsto U(g)\,B\,U(g)^{-1}$ is 
smooth in the norm topology. Buchholz and Epstein
consider a special class of spacelike cones: Let 
$\underline{\cone}\subset \Bb^3$ be an open, salient cone in the $x^0=0$ 
plane of Minkowski space, with apex at the origin. Then its causal completion 
$\cone=\underline{\cone}''$ is a spacelike cone. Its dual cone $\cone^*$ is defined
as the set 
\[  \cone^*\doteq\big\{\; (p^0,\bfp)\in\Bb^4\;:\, 
\bfp\cdot\bfx>0\;\mbox{  for all } 
\bfx\in\underline{\cone}^\clo \setminus\{0\}\; \big\} \,.
\]
\begin{Lem}[Buchholz, Epstein] \label{LemBuEp} 
Let $B\in\F_{\sec,k}^\infty(\cone),$ where $\cone$ is a spacelike cone as
above and $\sec\in\Sece.$  
Then $p\mapsto \big(\Ee\,B\,\Omega\big)(p)$ is,
considered as a function on the mass shell $H_\msec,$ the smooth boundary
value of an analytic function $k\mapsto \big(\Ee\,B\,\Omega\big)(k)$ 
on the simply connected subset 
\[ 
\Gamma_{\cone,\sec}\doteq\{ k\in\Bc^4\;|\; k^2=m_\sec^2\;,\; 
\im k\in-\cone^*\; \} 
\]
of the complex mass shell. Further,  its boundary value on $-H_\msec$
satisfies  
\begin{equation} \label{eqBuEp}
\omega_{\sec,k}\,V_{\ssec}(\frac{1}{\msec}\utilde{p} \sigma_2)\,
 \overline{\big(\Ee B\,\Omega\big)(-p)}
=\big(\Cop_{\sec,k}^*\Ee B^*\,\Omega\big)(p)\;,
\end{equation}
where $\omega_{\sec,k}$ is a complex number of unit modulus which is 
independent of $B$ and $\underline{\cone}$, and $\Cop_{\sec,k}^*$ is
the mentioned intertwiner from $\Ee\,\F_{\sec,k}(\cone)^*\,\Omega^\clo$ onto  
$\Eeseck\,\calH$. 
\end{Lem} 
Note that equation~\eqref{eqBuEp} coincides literally with equation~(5.13)
in~\cite{BuEp}. 
We reformulate this result as follows. Denote by $\calK_1$ the
class of spacelike cones $\cone$ contained in $W_1$ which are of the form 
$\underline{\cone}''$ as in the lemma and contain the positive $x^1$-axis. 
Let further 
\begin{align} \label{eqD0}
D_0\doteq
\underset{\cone\in\calK_1,\sec,k}{{\rm span}}\;
\Ee\,\F_{\sec,k}^\infty(\cone)\,\Omega 
\end{align}
where $\sec$ runs through $\Sece$ and $k=1,\ldots,d_\sec.$ 
The lemma asserts that on this domain an operator $S_0$ may be 
defined by 
\begin{align}   \label{eqS0}
 \big(S_0\Eeseck\psi\big)(p)\doteq 
V_{\ssec}(\frac{1}{\msec}\utilde{p}\sigma_2)\,
\overline{\big(\Eeseck\psi\big)(-p)}\,,\quad\psi\in D_0\,. 
\end{align}
Further, the intertwiners $\Cop_{\sec,k},$ modified by the factors 
$\omega_{\sec,k}$ appearing in~\eqref{eqBuEp}, extend by linearity to a 
unitary ``charge conjugation'' operator $\Cop$ on $\Ee\calH,$ 
\[ \Cop \,\Eeseck \doteq \omega_{\sec,k}\Cop_{\sec,k}\,\Eeseck\,,
\]
which satisfies the equations~\eqref{eqCop} of the proposition. 
Now equation~\eqref{eqBuEp} may  be rewritten as 
\begin{equation} \label{eqBuEp'}
 \Cop\,S_0\subset \Stome\,. 
\end{equation}
This inclusion implies in 
particular that $S_0$ is closable, its closure satisfying the same  relation.
But this closure is an extension of the operator $\Sgeoe,$ as we show
in the Appendix (Lemma~\ref{LemSgeoStom}). Hence we have 
\begin{equation}  \label{eqSgeoStom}
  \Cop\,\Sgeoe \subset \Stome\,,
\end{equation} 
and it remains to show the opposite inclusion. To this end, 
we refer to the opposite wedge $W_1'=R_2(\pi)\, W_1.$ Let 
\begin{align*}
\Sgeoe'&\doteq\Ue(r_2(\pi))\;\Sgeoe\;\Ue(r_2(\pi))^{-1}\,,\\
\Stome'&\doteq\Ue(r_2(\pi))\,\Stome\,\Ue(r_2(\pi))^{-1}\,
=S_{\rm Tom}(W_1')\,\Ee\,.
\end{align*}
We claim that the following sequence of relations holds true: 
\begin{equation} \label{eqRLLR}
\Stome\subset \kappa^{-1}\,(\Stome')^*
\subset\kappa^{-1}\,\Cop\,(\Sgeo')^*=\Cop\,\Sgeoe\;,
\end{equation}
where $\kappa$ is the Bose-Fermi operator. 
Twisted locality and modular theory imply that 
\[ Z\Stom Z^*\subset S_{\rm Tom}(W_1')^*. \] 
Applying $\Ee,$ this yields $Z\Stome Z^*\subset$ $ (\Stome')^*.$ 
But $Z\Stome Z^*=$ $Z^2\Stome,$ because 
$\kappa$ commutes with the modular operators. Using $Z^2=\kappa,$ 
this proves the first inclusion. 
Since $\Cop$ commutes with $\Ue(\Potild)$ and both $\Sgeo$ and
$\Stome$ are involutions, the inclusion~\eqref{eqSgeoStom} implies that 
\begin{equation} \label{eqCj}
  \Cop\Uej=\Uej\Cop^*\,. 
\end{equation} 
%To see this, let $\psi\in\dom \Sgeoe \cap \calH_\sec.$ Then by 
%equation~\eqref{eqSgeoStom}, $(\Stome)^2\psi=
%\overline{\omega_{\sec,k}}\,\omega_{\bar\sec}\,(\Sgeoe)^2\psi.$ Since 
%both $\Sgeoe$ and $\Stome$ are involutions, this shows that 
%that $\omega_{\bar\sec}=\omega_{\sec,k},$ and hence 
%$\omega\,\Sgeoe= \Sgeoe\,\omega^*.$ 
Thus, the adjoint of relation~\eqref{eqSgeoStom} reads 
$\Stome{^*}\subset \Cop\Sgeoe{^*},$ which implies the second 
of the above inclusions.  
Finally, the group relations~\eqref{eqjr2j} and 
$\lambda_1(t)\,r_2(\pi) = r_2(\pi) \,\lambda_1(-t)$ 
imply that $(\Sgeoe')^*=\Ue(r_2(2\pi))\,\Sgeoe.$ But the
spin-statistics theorem~\cite{BuEp} asserts that $U(r_2(2\pi))=\kappa.$ 
(Namely, both operators act on $\Hsec$ as multiplication by the
statistics sign $\kappa_\sec=e^{2\pi i  s_\sec}.$) 
Hence the last equation 
in \eqref{eqRLLR} holds. 
This completes the proof of \eqref{eqRLLR} and hence of the proposition. 
\end{proof} 

By uniqueness of the polar decomposition, equation~\eqref{eqCSS} of the 
proposition implies the equations 
\begin{equation*} 
 \DWR^{\half}\;\Ee= e ^{-\pi K}\;\Ee\,,
\quad \JWR\;\Ee  = \Cop\,\Uej\;\Ee \,.
\end{equation*}
Since the unitary $\Cop$ commutes with $\Ue(\Potild)$ and satisfies
equation~\eqref{eqCj}, we have shown the 
single particle version of the Bisognano-Wichmann theorem: 
\begin{Thm} \label{ThmModCov1} 
Let the assumptions 0)$,\ldots,$ vi) of Section~\ref{secAssRes} hold. Then 

\noindent i)  Modular Covariance holds on the single particle space: 
\begin{align}  \label{eqModCov1}
   \DWR^{it}\;\Ee&= U(\lambda_1(-2\pi t))\;\Ee\,. \\
\intertext{ii) $\JWR\;\Ee$ is a ``CPT operator'' on $\Ee\calH$:} 
   \JWR\,U(g)\JWR\;\Ee &=U(jgj) \;\Ee \quad\text{ for all } g\in\Potild\,. 
  \label{eqJgJ1}
\end{align}
\end{Thm}
\section{Modular Covariance on the Space of Scattering States.} \label{secHex} 
Having established modular covariance on the single particle space, we
now show that it extends to the space of scattering states. The
argument is an extension of Landau's analysis~\cite{Landau} on the
structure of local internal symmetries to the present case of a symmetry which
does not act strictly local in the sense of Landau. The method 
to be employed is Haag-Ruelle scattering theory~\cite{Hepp, Jost},
whose adaption to the present situation of topological charges has
been developed in~\cite{BuF}. 

This method associates a multi-particle state to $n$ 
single particle 
vectors, which are created from the vacuum by quasilocal field
operators carrying definite charge.  
Recall~\cite{DHRI} that for every $\sec\in\Sec,$
there is a family of linear subspaces $\cone \rightarrow
\F_\sec(\cone)\subset\F(\cone)$ of field 
operators carrying charge $\sec:$ 
\[ \F_\sec(\cone)\,\Omega=E_\sec\,\F(\cone)\,\Omega\,. 
\]
Operators in $\F_\sec(\cone)$ are bosons or fermions  w.r.t.\ the
normal commutation relations ac\-cording as $\kappa$ takes the value $1$ or
 $-1$ on $\Hsec.$  
The mentioned quasilocal creation operators  are constructed as follows. 
For $\sec\in \Sece,$ let $B\in\F_\sec(\cone)$ be such that the spectral 
support of $B\Omega$ has 
non--vanishing intersection with the mass hyperboloid $H_\msec.$ 
Further, let $f\in\calS(\Bb^4)$ be a Schwartz function whose Fourier transform 
$\tilde{f}$ has compact support contained in the open forward light cone $V_+$ 
and intersects the energy momentum spectrum of the sector $\sec$
only in the mass shell $H_\msec.$ 
Recall that the latter is assumed to be isolated from the rest of the
energy momentum spectrum in the sector $\calH_\sec.$ 
For $t\in\Bb,$ let $f_t$ be defined by  
\begin{equation}  \label{eqft}
f_t(x)\doteq (2\pi)^{-2}\,\int\d^4 p \:
e^{i(p_0-\omega_\sec(\bfp) t}\,e^{-ip\cdot x}\,\tilde{f}(p) \;, 
\end{equation}
where $\omega_\sec(\bfp)\doteq(\bfp^2+m_\sec^2)^{\half}.$ For large $|t|,$
its support is essentially contained in the region $t \,V_\sec(f),$
where $V_\sec(f)$ is the velocity support of $f,$ 
\begin{equation}  \label{eqVf}
V_\sec(f)\doteq \{\, \big(1,\frac{\bfp}{\omega_\sec(\bfp)}\big)\;,\;
p=(p^0,\bfp)\in\supp\tilde{f}\,\}\,.   
\end{equation}
More precisely~\cite{BBS,Hepp}, for any $\eps>0$ there is a Schwartz function 
${f}^\eps_t$ with support in $t\,V_\sec(f)^\eps,$ where $V^\eps$
denotes an $\eps$--neighbourhood of $V,$ such that
$f_t-{f}^\eps_t$ converges to zero in the Schwartz topology for 
$|t|\rightarrow\infty.$
Let now 
\begin{equation*} 
 B(f_t)\doteq\int \d^4 x \,f_t(x)\,U(x) B U(x)^{-1} \,.
\end{equation*}
For large $|t|,$ this operator is essentially localized 
in $\cone+t\,V_\sec(f).$ 
Namely, for any $\eps>0,$ it can be approximated by the operator 
\begin{equation} \label{eqBeps}
B(f^\eps_t)\in\F\big(\cone+t\,V_\sec(f)^\eps\big) 
\end{equation}
in the sense that  $\|B(f^\eps_t)-B(f_t)\|$ is of fast decrease in $t.$ 
Further, it creates from the vacuum a single particle vector 
\begin{equation*}  
B(f_t) \,\Omega=(2\pi)^2
\tilde{f}(P)\,B\,\Omega\quad\in\,\Hesec\,,
\end{equation*}
which is independent of $t,$ and whose velocity support is contained
in that of $f.$  Here we understand the velocity support $V(\psi)$ of
a single particle vector to be defined as in equation~\eqref{eqVf}, with
the spectral support of $\psi$ taking the role of $\supp \tilde{f}.$ 
To construct an outgoing scattering state from $n$  single
particle vectors, 
pick $n$ localization regions  $\cone_i,$ $i=1,\ldots,n$ 
and compact sets $V_i$ in
velocity space, such that for suitable open neighbourhoods
$V_i^\eps\subset \Bb^4$ the regions 
$\cone_i+t\,V_i^\eps$ are mutually spacelike separated for large $t.$ 
Next, choose $B_i\in\F_{\sec_i}(\cone_i),$ and Schwartz functions $f_i$ as
above with $V_{\sec_i}(f_i)\subset V_i.$ Then the standard lemma of
scattering theory asserts the following: The limit 
\begin{equation}  \label{eqBn1Om}
\lim_{t\rightarrow\infty} B_n(f_{n,t})\cdots B_1(f_{1,t})\,\Omega \doteq
\big(\psi_n\times\cdots\times\psi_1\big)^{{\rm out}}\, 
\end{equation}
exists and depends  only on the single particle vectors 
$\psi_i\doteq B_i(f_{i,t})\,\Omega,$ justifying the above notation. 
The convergence  in \eqref{eqBn1Om} is of fast decrease in $t,$ and
the limit vector depends continuously on the single particle
states, as a consequence of the cluster theorem.  
Further, the normal commutation relations survive in this limit. 
%It is noteworthy that these scattering states do not 
%depend~\cite{BuF} on the Lorentz frame chosen in
%\eqref{eqft}. In fact, there is an intrinsically Lorentz-independent 
%formulation~\cite{FGR} of scattering theory, but it is not suitable 
%for our purposes. 

Let us write $\He\doteq\Ee\calH,$ and denote by $\calH^{(n)},$
$n\geq2,$ the closed span of outgoing $n$-particle scattering states 
and by $\calH^{(\ex)}$ the  span of these spaces: 
\begin{equation} \label{eqHex}
\calH^{(\ex)}=\Bc\,\Omega\oplus\bigoplus_{n\in\Bn} \calH^{(n)}\,.
\end{equation}
Asymptotic completeness means that $\calH^{(\ex)}$ coincides 
with $\calH.$ 
Our proof that modular co\-va\-riance extends from $\He$ to
$\calH^{(\ex)}$ relies on the following observation. 
\begin{Lem} \label{Velo}
In each $\calH^{(n)},$ $n\geq 2,$ there is a total set of scattering states as 
in equation~\eqref{eqBn1Om}, with the localization regions chosen such
that $\cone_1,\dots,\cone_{n-1}\subset W_1'$ and $\cone_n= W_1.$  
\end{Lem}
In particular, for these scattering states the regions
$\cone_i+tV(\psi_i)^\eps,$ $i=1,\ldots,n-1,$ are spacelike 
separated from  $W_1+tV(\psi_n)^\eps$ for large $t.$ 

\begin{proof}
Consider the set $M^n$ of velocity tupels $(\bfv_1,\dots,\bfv_n)\in\Bb^{3n}$
satisfying the requirements that  a) one of the velocities, 
say $\bfv_{i_0},$ has the strictly largest 
$1$-component: 
\[ (\bfv_{i_0})_1 > (\bfv_{i})_1  \quad\text{ for } i\neq i_0\,,\]
and b) the relative velocities w.r.t. $\bfv_{i_0}$ have different directions: 
\[ \Bb^+\,(\bfv_i-\bfv_{i_0}) \neq \Bb^+\,(\bfv_j-\bfv_{i_0}) 
\quad\text{ for } i\neq j \,.\]
Given such $(\bfv_1,\ldots,\bfv_n),$ let $\cone_{i_0}= W_1.$ For $i\neq
i_0,$ let $\underline{\cone}_i$ be a cone  in the $t=0$ plane of $\Bb^4$ 
containing the ray $\Bb^+\,(\bfv_i-\bfv_{i_0})$ and with 
apex at the origin, and let then $\cone_i$ be its causal closure. 
Then, having chosen sufficiently small
opening angles, the regions $\cone_i+t\{(1,\bfv_i)\},$ $i=1,\ldots,n,$ are  
mutually spacelike separated for all $t>0,$ and
further $\cone_i\subset W_1'$ for $i\neq i_0.$  
Now $M^n$ exhausts the set of all velocity tupels in $\Bb^{3n}$ except
for a set of measure
zero. Hence, a scattering state $(\psi_n\times\cdots\times\psi_1)^{{\rm out}}$ 
as in \eqref{eqBn1Om} can be approximated by a sum of scattering states
$(\psi^\nu_n\times\cdots\times\psi^\nu_1)^{{\rm out}},$ whose localization 
regions satisfy that $\cone^\nu_{i_0}= W_1$ for some $i_0,$ and 
$\cone^\nu_i\subset W_1'$ for $i\neq i_0.$ This is accomplished by a standard 
argument~\cite{Araki} taking into account the continuous dependence of 
$(\psi_n\times\cdots\times\psi_1)^{{\rm out}}$ on the $\psi_i$ and the 
Reeh-Schlieder theorem. 
But due to the normal commutation relations obeyed by the scattering
states, %the $i_0^{\rm th}$ factor may be permuted to the left, hence 
$(\psi^\nu_n\times\cdots\times\psi^\nu_1)^{{\rm out}}$ coincides with $\pm\,
(\psi^\nu_{i_0}\times\cdots\psi^\nu_n\cdots\times\psi^\nu_1)^{{\rm out}}$ 
and hence is of the form required in the lemma. 
\end{proof}
\begin{Prop} \label{ModCovScat}
 If the unitary groups $\DWR^{it}$ and $U(\lambda_1(-2\pi t))$
 coincide on $\He,$ they also coincide on the space $\calH^{(\ex)}$ of
 scattering states. 
\end{Prop}
\begin{proof}
Let $U_t\doteq  \DWR^{it}\,U(\lambda_1(2\pi t)).$ Considering this
operator  as an internal symmetry, it should act multiplicatively on 
the scattering states as shown by Landau in~\cite{Landau}. The
complication is that $U_t$ does not act strictly local, but only leaves
$\F(W_1)$ invariant. We generalize Landau's argument to this case
utilizing the last lemma. By induction over the particle
number $n$ we show that 
$U_t$ is the unit operator on each $\calH^{(n)}.$ 
Let $(\psi_n\times\cdots\times\psi_1)^{{\rm out}}$ be a scattering
state with $\psi_i=B_i(f_{i,t}),$ where the localization regions $\cone_i$ are
as in the above lemma. Since
$\|B_{n-1}(f_{n-1,t})\cdots B_{1}(f_{t})\Omega-
(\psi_{n-1}\times\cdots\times\psi_1)^{{\rm out}}\|$ is of fast
decrease in $t,$ while $\|B_n(f_{n,t})\|$ increases at most like $|t|^4,$ one
concludes as Hepp in~\cite{Hepp}: 
\begin{align} \label{eqn(n-11)}
(\psi_n\times\cdots\times\psi_1)^{{\rm out}}&=\lim_{s\to\infty} 
B_n(f_{n,s})\,(\psi_{n-1}\times\cdots\times\psi_1)^{{\rm out}} \,.
\intertext{ Hence }
U_t\,(\psi_n\times\cdots\times\psi_1)^{{\rm out}}&=\lim_{s\to\infty} 
U_tB_n(f_{n,s})U_t^{-1}\,(\psi_{n-1}\times\cdots\times\psi_1)^{{\rm out}}\;,
\label{eqUtBpsi}
\end{align}
where we have put in the induction hypothesis that $U_t$ acts trivially on 
$\calH^{(n-1)}.$  
Due to  Borchers' result, $U_t$ commutes with the translations, which
implies that $U_tB_n(f_{n,s})U_t^{-1}$ coincides with 
$(U_tB_nU_t^{-1})(f_{n,s}).$ But modular theory and covariance
guarantee that $U_tB_nU_t^{-1}$ is, like $B_n,$ in $\F(W_1).$ In
addition $(U_tB_nU_t^{-1})(f_{n,s})$ $\Omega=$ $U_t\psi_n,$ hence by
the standard lemma of scattering theory, equation~\eqref{eqUtBpsi}
may be rewritten as 
\[U_t\,(\psi_n\times\cdots\times\psi_1)^{\rm out}= 
((U_t\psi_n)\times\psi_{n-1}\times\cdots\times\psi_1)^{\rm out}\,.
\]
By assumption of the proposition, $U_t$ acts trivially on $\psi_n,$ and 
hence on the scattering state. By linearity  and continuity, the
same holds on $\calH^{(n)},$ completing the induction. 
\end{proof}
The hypothesis of this proposition has been shown in
Theorem~\ref{ThmModCov1} to hold under our assumptions 0)$,\ldots,$
vi). Hence, we have now derived modular covariance  from these
assumptions and asymptotic completeness. 
As mentioned, Guido and Longo have shown that modular covariance 
generally implies covariance of the modular conjugations, and hence
the CPT theorem~\cite[Prop.~2.8,~2.9]{GL}. 
Thus, the proof of Theorem~\ref{BiWi} is now completed. 
\section{The CPT Theorem.} \label{secCPT}
We show here that the CPT theorem can also be derived directly   
from our assumptions in Section~\ref{secAssRes}, 
via the single particle result and scattering theory. This should in
particular turn out useful for a derivation of the CPT theorem in a 
theory of massive particles with non--Abelian braid group statistics
(plektons) in $d=2+1,$ where the methods of~\cite{GL} cannot be
applied in an obvious way since one has no field algebra. 

Recall that incoming scattering states can be constructed as in
equation~\eqref{eqBn1Om}, where now $t\rightarrow -\infty$ and the
condition for the limit to exist is that the regions 
$\cone_i-|t|\,V_i^\eps$ be 
mutually spacelike separated for large $|t|$. The following
result 
holds under the assumptions of Section~\ref{secAssRes}, but without 
restrictions on the degeneracies of the mass eigenvalues in each
sector. 
Like Proposition~\ref{ModCovScat}, it is an extension of Landau's
argument~\cite{Landau}. 
\begin{Lem} \label{JScat}
$\JWR$ maps outgoing scattering states to incoming ones and vice versa
according to 
\begin{equation} \label{eqJScat}
\JWR\,\big(\psi_n\times\cdots\times\psi_1\big)^{\Out}=
\big(\JWR\psi_n\times\cdots\times\JWR\psi_1\big)^{\In} \,. 
\end{equation}
\end{Lem}
Let us put the statement of the lemma into a more concise form. 
Recall that the spaces of incoming and outgoing scattering
states are isomorphic to an appropriately symmetrized Fock space over
$\He$ via the operators $W_{\In,\Out}$ which map 
$\psi_n\otimes\cdots\otimes\psi_1$ to 
$(\psi_n\times\cdots\times\psi_1)^{\In,\Out},$ respectively. 
 Lemma~\ref{JScat} then asserts that 
\begin{equation}  \label{eqJS}
 \JWR\,W_{\Out}=W_{\In}\, \Gamma(\JWR\Ee)\,,
\end{equation}
where $\Gamma(U)$ denotes the second quantization of a unitary operator $U$
on $\He.$  
Note that the same equation holds with $W_ {\Out}$ and
$W_{\In}$ interchanged. 

\begin{proof}
We proceed by induction along the same lines as in the last
proposition. Let $\psi_i=B_i(f_{i,t})\Om$  be the single particle states 
appearing in equation~\eqref{eqJScat}, 
with velocity supports contained in
compact sets $V_i,$ and with localization regions $\cone_i,$ such that 
$\cone_i+tV_i^\eps$ are mutually spacelike separated for suitable $\eps>0.$   
According to Lemma~\ref{Velo}, we may assume that the localization
regions satisfy $\cone_1,\dots,\cone_{n-1}\subset W_1'$ and $\cone_n= W_1.$ 
By the same arguments as in the last proof, we have 
\begin{align}  
\JWR\;(\psi_n\times\cdots\times\psi_1)^{{\rm out}}&=
\lim_{t\to\infty} \JWR\,B_n(f_{n,t})\,\JWR\;
(\JWR\psi_{n-1}\times\cdots\times\JWR\psi_1)^{{\In}} \;, \label{eqJpsiin}
\end{align}
where we have put in the induction hypothesis that $\JWR$ acts as in 
equation~\eqref{eqJScat} on $\calH^{(n-1)}.$ Now by Borchers'
result~\cite{Borch92}  
we know that the  commutation relations $\JWR\,U(x)\JWR$ $=U(jx)$ hold. 
From these we conclude that the spectral supports of
$\psi\in\calH$ and $\JWR\psi$ are related by the transformation $-j,$ and
hence their velocity supports are related by 
\begin{equation} \label{eqVJ} 
V(\JWR\psi)=  -r\,V(\psi) \;, 
\end{equation}
where $r$ denotes the inversion of the sign of the $x^1$-coordinate.  
By virtue of the Reeh-Schlieder theorem and the continuity of the
scattering states, we may assume that for $i=1,\ldots,n-1$ 
there are $\hat{B}_i\in\F(r\,\cone_i)$ and $\hat{f}_i$ such that 
$\hat{B}_i(\hat{f}_{i,-t})\Omega=\JWR\psi_i.$ Further, $\hat{f}_i$ can
be chosen such that $V(\hat{f}_i)\subset V(\JWR\psi_i)^\eps,$ which in
turn is contained in  $-r\,V_i^\eps$ due to equation~\eqref{eqVJ}.
Then $\hat{B}_i(\hat{f}_{i,-t})$ can be approximated by an operator 
$A^\eps_i(t)$ localized in the region $r\,\{\cone_i+tV_i^\eps\}.$ 
These regions are mutually spacelike separated for large positive 
$t,$ and hence the incoming 
$n-1$ particle state in equation~\eqref{eqJpsiin} may be  written
as $\lim_{t\rightarrow-\infty}$ 
$\hat{B}_{n-1}(\hat{f}_{n-1,t})\cdots \hat{B}_1(\hat{f}_{1,t})\,\Omega.$ 
Similarly, Borchers' commutation relations imply that 
\[  
\JWR\,B_n(f_{n,t})\,\JWR =
\big(\JWR B_n \JWR\big)\,(\hat{f}_{n,-t})\;,
\quad\text{  where } \,\,\hat{f}_n(x)=\overline{f_n(jx)}\,.
\]
Now $\JWR\,B_n\,\JWR$ is in $\F(W_1)',$ and $V(\hat{f}_n)=-r\,V(f_n),$ 
and therefore the discussion around 
equation~\eqref{eqBeps} implies that the above operator may be
approximated by an operator $A^\eps_n(t)\in\F\big(W_1+t\cdot rV_n^\eps\big)'.$ 
Recall that the operators $A^\eps_i(t),$ $i=1,\ldots,n-1,$ are localized
in the regions $r\,\{\cone_i+tV_i^\eps\}.$ 
For large positive $t,$ these regions are spacelike to $r\,\{W_1+tV_n^\eps\}$ 
and are hence contained in $W_1+t\cdot rV_n^\eps.$ 
Hence the $A^\eps_i(t)$ (anti-) commute with $A^\eps_n(t)$ for large $t.$ 
Thus the standard arguments of scattering 
theory~\cite{Hepp,DHRIV} apply, yielding that the vector 
\eqref{eqJpsiin} may be written as 
\begin{align*}
\lim_{t\to-\infty} \big(\JWR B_n \JWR\big)(\hat{f}_{n,t}) 
\;\hat{B}_{n-1}(\hat{f}_{n-1,t})\cdots\hat{B}_{1}(\hat{f}_{1,t})\,\Omega\;,
%=  \lim_{t\to+\infty}A^\eps_n(t)\cdots A^\eps_1(t)\,\Omega\;, 
\end{align*}
and only depends on the single particle vectors. 
But these are $\big(\JWR B_n \JWR\big)$ $(\hat{f}_{n,t})$ 
$\Omega=\JWR\psi_{n},$
and $\hat{B}_{i}(\hat{f}_{i,t})\,\Omega=\JWR\psi$ for $i=1,\ldots,n-1.$ 
Hence the limit coincides with the right hand side of 
equation~\eqref{eqJScat}, completing the induction. 
\end{proof}
\begin{Prop}[CPT] \label{CPT} 
Let $\CPTop$ be the the anti--unitary involution $\CPTop\doteq Z^*\,\JWR.$ 

\noindent i)If  the representation property 
\begin{equation} \label{eqJgJ}
 \CPTop\,U(g)\,\CPTop= U(jgj)\quad\text{ for all } g\in\Potild 
\end{equation}
holds on $\He,$ then it is also satisfied on the space $\calH^{(\ex)}$ of 
scattering states. 

\noindent ii) In this case, and if in addition asymptotic completeness holds, 
$\CPTop$ acts geometrically correctly on the family of wedge
algebras $\F(W)_W$ in the sense of equation~\eqref{eqjWj}. 
\end{Prop}
Note that equation~\eqref{eqJgJ} is equivalent to $\JWR U(g)\JWR=U(jgj),$ 
since $Z$ commutes with $U(g)$ and satisfies
$Z^*\JWR=\JWR Z.$ In Proposition~\ref{ThmModCov1}, we have shown that 
$\JWR$ satisfies this representation property on $\He$ if the
assumptions 0)$,\ldots,$vi) of Section~\ref{secAssRes} hold. Hence
Proposition~\ref{CPT} is a CPT theorem, holding under these assumptions and
asymptotic completeness. 

\begin{proof}
i) Let $\JWR$ have the above  representation property on $\He.$ As is well  
known~\cite{DHRIV}, the restriction of $U(\Potild)$
to the space of scattering states is equivalent to the second 
quantization of its restriction to $\He:\,$ 
$U(g)\,W_{\Out,\In}$ $=W_{\Out,\In}\,$ $\Gamma(U(g)\Ee).$  By virtue of 
Lemma~\ref{JScat}, see equation~\eqref{eqJS}, the assumption thus implies 
\begin{align*}
\JWR U(g)\JWR\;W_{\Out}&=W_{\Out}\;\Gamma\big(\JWR U(g)\JWR\Ee\big) =
% W_{\Out}\;\Gamma\big(U(jgj)\Ee\big) = 
U(jgj)\;W_{\Out}\;, 
\end{align*}
which proves the claim. 
ii) By twisted locality and modular theory, one has 
\begin{equation} \label{eqjWRj} 
\F(W_1') \subset Z^*\,\F(W_1)'\,Z=\CPTop\,\F(W_1)\,\CPTop\,.
\end{equation}
Now recall that %$W_1'=R_2(\pi)\cdot W_1,$ hence 
$U(r_2(\pi))\,\F(W_1)\,U(r_2(\pi))^{-1}=\F(W_1')$  and that
$jr_2(\pi)j=r_2(-\pi),$ see equation~\eqref{eqjr2j}. 
One therefore obtains, by applying
${\rm Ad}\big(U(r_2(\pi))\CPTop\big)$ to the inclusion~\eqref{eqjWRj} 
and using equation~\eqref{eqJgJ}, the opposite inclusion. 
Hence equality holds in~\eqref{eqjWRj}. Since every
wedge region arises from $W_1$ by a Poincar\'e transformation, 
the claimed equation~\eqref{eqjWj} follows by covariance of the field
algebras and the representation property~\eqref{eqJgJ} of $\CPTop.$ 
\end{proof}
%%%%%%%%%%%%%%%%%%%%%%%%%%%%%%%%%%%%%%%%%%%%%%%%%%%%%%%%%%%%%%%%%%%% !!!!!!
\appendix

\section{Single-Particle PT Operator and Geo\-metric \\ In\-volution.} 
We provide an explicit formula for the group relations $jgj$ and a
proof of the representation property of the ``PT operator'' 
$\Ums(j)$ defined in equation~\eqref{eqU1j}. 
As before, we denote by $g\mapsto jgj$ the unique lift~\cite{Var2} of 
the adjoint  action of $j$ on the  Poincar\'e group to an automorphism
of the covering group. An explicit formula for $jgj$ follows from the 
observation that $j$ coincides with the proper Lorentz transformation 
$-R_1(\pi):$ Hence, for all $A\in SL(2,\Bc)$  
\[  j\,\Lambda(A) \,j=R_1(\pi)\,\Lambda(A) \,R_1(\pi)^{-1}
=\Lambda(\sigma_1\,A\,\sigma_1)\,.
\] 
This shows that the lift $jgj$ is given by 
\begin{equation}  \label{eqjgj}
j\,(x,A)\,j=(j\, x,\, \,\sigma_1\,A\,\sigma_1) \quad \text{for all }
(x,A)\in\Potild\,.
\end{equation} 
Using equation~\eqref{eqBooRot}, one has in particular the relations 
\begin{equation} \label{eqjr2j} 
j\,r_2(\omega)\,j= r_2(-\omega)\,,\quad 
j\,\lambda_1(t)\,j= \lambda_1(t)\,.
\end{equation}
\begin{Lem} \label{LemU1j} 
The operator $\Ums(j)$ defined in equation~\eqref{eqU1j} is
anti--unitary and satisfies the representation properties~\eqref{eqUjgj1}. 
\end{Lem}
\begin{proof}
We prove the second of the equations~\eqref{eqUjgj1} for 
$g=(0,A)$ with $A\in SL(2,\Bc)$. The other assertions are shown along 
the same lines. Recall that the covering homomorphism 
$\Lambda:SL(2,\Bc)\rightarrow\Lor$ is characterized by \vspace{-10pt} 
$ \utilde{\Lambda(A)\, p}=A\,\utilde{p}\;A^*.$  
We have 
\begin{multline} \label{eq1Ujgj}
\big(\Ums(j)\,\Ums(g)\,\Ums(j)\,\psi\big)(p)\\ 
=V_s\big(m^{-2}\utilde{p}\,\sigma_3\,\bar{A}\;
  \overline{\utilde{(\Lambda(A^{-1})(-j)p)}}\,
         \sigma_3 \big)\,\psi(j\Lambda(A^{-1})jp)\,. 
\end{multline} %\vspace{-10pt}
Using the identity 
\[
\utilde{\Lambda(A^{-1})(-j)p}=\utilde{\Lambda(A^{-1}i\sigma_1)p}
=A^{-1}\sigma_1 \utilde{p} \sigma_1 (A^* )^{-1}\;,  \vspace{-15pt}
\]
which follows from  $-j=R_1(\pi),$ and the well-known relation 
\begin{align*}
\sigma_2 \bar{A} \sigma_2=(A^* )^{-1}  % \quad\text{ and }\quad 
% \sigma_1 (\bar{A}^* )^{-1}\sigma_1 =\sigma_3 A \sigma_3
\quad\text{  for } A\in SL(2,\Bc)\;,
\end{align*}
one verifies that the argument of $V_s$ in equation~\eqref{eq1Ujgj} equals 
$\sigma_1\,A\,\sigma_1.$ By equation~\eqref{eqjgj}, this proves the claim. 
\end{proof}

We now relate the geometric involution $\Sgeoe=\Uej e^{-\pi K}\Ee$
with the closable operator $S_0$ defined in equation~\eqref{eqS0}. 
\begin{Lem} \label{LemSgeoStom} 
The closure of  $S_0$ is an extension of $\Sgeoe.$ 
\end{Lem}
\begin{proof}
Recall that for $f\in\calS(\Bb)$ the bounded operator $f(K),$ where
$K$ denotes again the generator of the boosts $\lambda_1(\cdot),$  
may be written as 
\[ f(K)=\int\d t\,\tilde{f}(t)\,U(\lambda_1(t)) \,. 
\] 
Here $\sqrt{2\pi}\tilde{f}$ is the Fourier transform of $f,$ and the integral is
understood in the weak sense.  
Let now $c$ be a smooth function with compact support, and let 
$\psi=\Ee B\Omega,$ where  $B\in\F_{\sec,k}^\infty(\cone)$
for some $\cone\in\calK_1.$ 
Applying the above formula to $c_\pi(K)\doteq e^{-\pi  K}c(K)$ 
one finds, using that $\tilde{c}$ is analytic and 
$\widetilde{c_\pi}(t)=\tilde{c}(t-i\pi),$ 
\begin{align}
\big(\Sgeoe\, c(K)\, \psi \,\big)(p)&= 
\int\d t\,\overline{\tilde{c}(t-i\pi)}
\big(\Uej \,U\big(\lambda_1(t)\big)\,\psi \,\big)(p) \label{eqJLPsi} \\ 
&=\int\d t\,\overline{\tilde{c}(t-i\pi)}\, 
V_\ssec\big(\frac{1}{\msec}\utilde{p} 
\sigma_3\,\overline{e^{\frac{t}{2}\sigma_1}}\big)\, 
 \overline{\psi\big(\Lambda_1(-t)(-jp)\big)}\,. \label{eqJLPsi'}
\end{align}
The one-parameter group $\Lambda_1(\cdot)$ extends to an
entire analytic function satisfying 
\[ \Lambda_1(-t-it') = \Lambda_1(-t)\big(j_{t'}-i \sin t'\; \sigma \big)\;,
\]
where $j_{t'}$ 
acts as multiplication by $\cos t'$ on the coordinates $x^0$ and
$x^1$ and leaves the other coordinates unchanged, and $\sigma$ acts as
$\sigma_1$ on $(x^0,x^1)$ and as the zero projection on
$(x^2,x^3)$~\cite{H96}. 
Note that in particular 
\begin{equation*} 
\Lambda_1(-t-i\pi)= \Lambda_1(-t) \,j\,.
\end{equation*}
Further, one easily verifies that for any $q\in H_\msec,$ the vector 
$\sigma\,q$ is in the dual cone $\cone^*.$ Hence for all $t'\in (0,\pi)$ 
and all $p\in H_\msec,$ the complex vector 
$\Lambda_1(-t-it')(-j p)$ is in $\Gamma_{\cone,\sec},$ the domain of
analyticity of $\psi,$ and approaches
$\Lambda_1(-t)(-p)$ as $t'\rightarrow\pi.$ 
It follows that the integrand in the expression~\eqref{eqJLPsi'} is
anti--holomorphic in $t$ in the strip $0<\im t<\pi,$ 
and that \eqref{eqJLPsi'} coincides with 
\begin{align} \nonumber 
&\int\d t\,\overline{\tilde{c}(t)}\, 
V_\ssec\big(\frac{1}{\msec}\utilde{p} 
\sigma_3\frac{1}{i}\sigma_1\,\overline{e^{\frac{t}{2}\sigma_1}}\big)\, 
 \overline{\psi(\Lambda_1(-t)(-p))}\;\\
&= \int\d t\,\overline{\tilde{c}(t)}\, 
\big(S_0U\big(\lambda_1(t)\big)\psi \,\big)(p) \,.\label{eqSLPsi}
\end{align}
Here we have used that for all $t,$ $U(\lambda_1(t))\,\psi$ is again in the 
domain $D_0$ of $S_0$ due to the covariance of the field 
algebra. This is so because for all $t,$ there is some 
$\cone_t\in\calK_1$ such that $\Lambda_1(t)\, \cone\subset \cone_t.$ 
Let now $\phi$ be in the (dense) domain of $S_0^*,$ and let $\psi\in D_0.$  
We have shown from \eqref{eqJLPsi} to \eqref{eqSLPsi}, that 
\begin{align*}
\big(\,\phi\,,\,\Sgeoe\, c(K) \psi \,\big)= 
\int\d t\,\overline{\tilde{c}(t)}
\big(\,\phi\,,\,S_0\,U\big(\lambda_1(t)\big)\,\psi \,\big) 
%&= \int\d t\,\overline{\tilde{c}(t)}
%\big(\,U\big(\lambda_1(t)\big)\,\psi \,,\,S_0^*\,\phi\,\big)\\
 = \big(\, c(K) \psi \,,\,S_0^*\phi\,\big)\,. 
\end{align*}
Let $D$ denote the set of finite linear combinations of 
vectors of the form $c(K)\,\psi,$ where $c\in C_0^\infty(\Bb)$ and
$\psi\in D_0.$ Then the above equation shows that $D$ 
is in the domain of $S_0^{**},$ %which is the closure of $S_0,$ 
and that $S_0^{**}=\Sgeoe$ on $D.$   
But $D$ is a core for $\Sgeoe,$ hence $S_0^{**}$ is an extension of $\Sgeoe.$ 
\end{proof}

\section*{Acknowledgements.} 
I thank K.~Fredenhagen, R.~Longo and D.~Buchholz for stimulating 
discussions which have been essential to this work, 
and K.-H.~Rehren for carefully reading the manuscript. 
Further, I gratefully acknowledge the hospitality extended to me by 
the Universities of Rome~I and II. 
Last not least, I acknowledge financial support by the SFB 288 
(Berlin), the EU (via TMR networks in Rome), the Graduiertenkolleg 
``Theoretische  Elementarteilchenphysik'' (Hamburg), and the DFG
(G\"ottingen).  

\providecommand{\bysame}{\leavevmode\hbox to3em{\hrulefill}\thinspace}

\end{document}